\documentclass[twocolumn,aps,amsmath,showpacs]{revtex4}  
\usepackage{graphicx,natbib}

\begin{document}

\title{Focal adhesion: Physics of a Biological Mechano-Sensor}
\author{Thomas Bickel and Robijn Bruinsma}
\affiliation{Department of Physics and Astronomy, UCLA, Box 951547, Los Angeles CA 90095-1547, USA}
\date{\today}

\begin{abstract}

\noindent Mechanical coupling between a cell and substrate relies on \emph{focal adhesions}, clusters of adhesion proteins linking stress fibers (bundles of actin proteins) inside the cell with surrounding tissue. Focal adhesions have been demonstrated to both \emph{measure} and \emph{regulate} the mechanical traction along the stress fibers. We present a quantitative model for focal adhesion mechano-sensing and stress regulation based on \emph{stress amplification} at the critical point of a condensation transition of the adhesion proteins.

\end{abstract}


\maketitle

Motor proteins transforming chemical energy into mechanical forces have become a textbook staple of biological physics~\cite{howardbook}. Less familiar to the physics community is the recent progress in our understanding of a molecular device, the \emph{focal adhesion}, that transmits, measures, and modulates mechanical force instead of generating it~\cite{geigerCOCB01,wehrleTCB02}. Focal adhesions form the mechanical links connecting the cytoskeleton to the extra-cellular matrix (ECM) that fills the space between cells in a living organism. They transmit tension forces generated inside the cell to the surrounding tissue, which can lead to cell locomotion~\cite{braybook}. Focal adhesions also play the role of \emph{tactile organs} for a cell: Assembly of a focal adhesion from an initial nucleus (the \emph{focal complex}) is partly determined by the \emph{rigidity} of the substrate, as well as by external forces applied to the cell. A mature focal adhesion also is an active, bi-directional source of signaling in and out of the cell that influences cell development.

\begin{figure}
\centering
\includegraphics{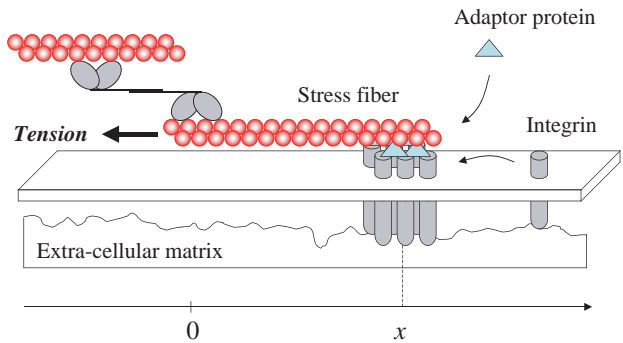} 
\caption{Structure of the focal adhesion, adapted form Riveline \emph{et al.}~\cite{rivelineJCB01}. Traction is generated along the action bundle (stress fiber) through myosin contraction. In response to the substrate compliance, a cluster of integrin and adaptor protein growths from an initial nucleus connected to the stress fiber.}
\label{fig1}
\end{figure}

The basic structure of the complex is shown in Fig.~\ref{fig1}.  An actin protein bundle, known as the ``stress fiber'', is connected via adaptor proteins (e.g., vinculin) to a highly elongated cluster of trans-membrane receptor proteins belonging to the \emph{integrin} family~\cite{geigerCOCB01}. Integrins bind reversibly to certain molecules of the ECM network (such as fibronectin or collagen), establishing adhesion between a cell and  substrate~\cite{joannyPRL03}. Mechanical tension is generated along the stress fiber through conventional actin-myosin II contractility (myosin II is the standard motor protein of muscles). Cell locomotion proceeds by formation of focal adhesions at the leading edge of the cell and generation of a sufficient level mechanical traction. Though cell motility has been extensively studied for a long time, it was discovered only recently that the traction level is determined by the \emph{rigidity} of the surrounding medium: On highly deformable substrates, the stress fiber tension level is low, focal adhesions are small and slip along the surface while on rigid substrates, focal adhesion are larger and the tension level is higher~\cite{pelhamPNAS97}. As a result, cells tend to crawl towards regions of increased substrate stiffness, which is known as \emph{durotaxis}~\cite{beningoTCB02}.

Elegant biophysical studies replacing the ECM by coated micron-sized beads~\cite{choquetCell97}, arrays of flexible microneedles~\cite{tanPNAS03}, or micropatterned substrates~\cite{loBPJ00,balabanNCB01,schwartzBPJ02}, revealed that, despite the structural and functional complexity of focal adhesions, their \emph{mechanical properties} are characterized by certain general features: 

\noindent (i) Tension: The maximum traction level $T$ of a mature, stationary focal adhesion is proportional to its \emph{surface area} with a stress level $\sigma$ of about 5 nN/$\mu \text{m}^2$~\cite{tanPNAS03,balabanNCB01}. 

\noindent (ii) Clutch Effect~\cite{smilenovSci98}: Micron-size beads covered with fibronectin (an ECM component) placed on a crawling cell near the leading edge are dragged along by the retrograde flow of actin polymers from the leading edge to the cell body. Like a``slipping clutch'', the bead initially can be restrained by the weak picoNewton level forces of an optical trap, but, after a latency period of about 10mn, the clutch``engages'' and an abrupt increase in the bead-restraining tension is observed pulling the bead out of the trap~\cite{choquetCell97,suterJCB98}.   

\noindent (iii) Mechano-Sensing: Applied force is the mechanical ``signal'' for the maturation of a focal complex to a focal adhesion.  This force need not be tension along the stress fiber but can as well be an externally applied force. In the latter case, growth of a focal adhesion can be mechanically stimulated \emph{even when myosin II activity is inhibited}~\cite{rivelineJCB01} . 

The mechanism behind focal adhesion traction regulation is currently not known. The relatively fast mechanical response times indicate that the mechanical switch involves \emph{a tension-induced conformational change} of one or more of the adaptor proteins~\cite{chicurelCOCB98}. This Letter proposes a model for mechano-sensing based on the concept that tension-induced conformational changes of the adaptor proteins modulate the critical point of a (two-dimensional) \emph{condensation transition} of integrins proteins. Integrins imbedded in reconsituted lipid vesicles have been shown to spontaneously condense into micron-size clusters when the vesicle is placed in contact with an adhesive substrate~\cite{guttenbergLangmuir00}. These ``primitive'' integrin clusters are however mechanically fragile, their size rapidly decreases under picoNewton level applied forces~\cite{komuraEPJE00}, and they do not exhibit any mechano-sensing properties, which requires the presence of the adaptor proteins linking the integrins together and connecting them to stress-fibers.

To illustrate the proposed mechanism, we use a simple model describing an (elongated) focal adhesion as a linear, one-dimensional cluster of $N$ identical integrin-adaptor (IA) units. Each unit is linked to a pair of helical actin filaments, as shown in Fig.~\ref{fig1}, and also is in contact with a substrate via a link that can slip. Let   $x(t)$ denote the position of the cluster along the substrate and let $v$  be the velocity of the actin bundle produced by myosin II activity. In the limit of low tensions, the total force $T$  exerted by the stress fiber on the link must be proportional to the slip rate
\begin{equation}
\label{fiber/cytoskeleton}
T=\Gamma_N (v-\dot{x})  \ .
\end{equation}
Since by assumption the cluster consists of $N$ identical units, the friction coefficient is proportional to $N$ so that  $\Gamma_N=N \gamma$ and the force f per IA unit equals $f=\gamma(v-\dot{x})$. Next, ECMÕs are in general non-Newtonian, visco-elastic media~\cite{braybook}  but we will describe -- again for simplicity -- the ECM as a simple Newtonian fluid with a (high) viscosity $\eta$. The substrate viscosity is the mechanical ``input signal'' that should determine cluster size and tension. The force required to drag a cylindrical rod of $N$ units through a viscous fluid with a velocity $\dot{x}$  equals
\begin{equation}
\label{fiber/substrate}
T=N\zeta \dot{x}  \ ,
\end{equation}
with $\zeta$ the unit substrate friction coefficient. For Newtonian fluids  $\zeta=2\pi\eta a/\ln(Na/r)$~\cite{doibook},   where $r$ is the rod radius and $a$ the length of one unit. It follows from Eqs.~(\ref{fiber/cytoskeleton}) and (\ref{fiber/substrate}) that $f=v(\gamma^{-1}+\zeta^{-1})^{-1}$  after elimination of  $\dot{x}$. Finally, the traction along the stress fiber must be proportional both to the number $N$ of actin-myosin filaments attached to the focal adhesion and to the velocity-dependent force exerted by individual myosin II motors. For purposes of discussion, we linearize the measured force-velocity curves~\cite{howardbook} 
\begin{equation}
\label{force/velocity}
T=\alpha N f_0 \left( 1-\frac{v}{v_0}\right)  \ .
\end{equation}
Here, $f_0$ is the myosin stall-force, $v_0$  the maximum stepping velocity, and $\alpha$  a dimensionless measure of the duty ratio of the motor activity and the number of motors per filament. The stall force $f_0$ and the characteristic velocity  $v_0$ define the natural scale $\zeta_0=f_0/v_0$  for the friction coefficients $\gamma$ and $\zeta$. Eq.~(\ref{force/velocity}) is consistent with Eqs.~(\ref{fiber/cytoskeleton}) and (\ref{fiber/substrate}) provided the tension per unit  equals
\begin{equation}
\label{force/friction}
f(\zeta)=f_0\frac{\zeta/\zeta_0}{1+k\zeta/\zeta_0}  \ ,
\end{equation}
with  $k=\alpha^{-1}+\zeta_0/\gamma$.

Eq.~(\ref{force/friction}) is a purely mechanical self-consistency condition that shows that stress-fiber tension increases with substrate viscosity, independently of the size of the focal adhesion. It provides us with a mechanical ``input signal'' for conformational changes of other proteins of the focal adhesion. However, when we estimate $f$ by equating it to the known stress $\sigma\approx 5\text{nM}/\mu\text{m}^2$  on a focal adhesion (see (i)) times the cross-section area of a single filament in an actin bundle (of order $10^4\text{\AA}^2$) we obtain a value of the order of 0.1 pN, comparable to the force level generated on segments of a flexible protein by \emph{thermal noise}. The unit tension is too weak to provide effective stress-detection.

According to (i)-(iii), stress-detection involves changes in \emph{size} of the focal adhesion. Assume that the focal adhesion is in thermal equilibrium with a two-dimensional solution of integrin proteins having a chemical potential $\mu=k_BT\ln c_0$. The cluster size statistics $P(N)$ for an $N$ protein cluster in thermal equilibrium is given by the Boltzmann distribution
\begin{equation}
\label{boltzmanncluster}
P(N)\propto c_0^Ne^{-\beta N \epsilon}  \ ,
\end{equation}
with  $\beta =(k_BT)^{-1}$ and $\epsilon$   the Helmholtz free energy per IA unit. A \emph{condensation transition} takes place when $\mu$ equals $\epsilon$. In order to describe the adaptor conformational transition, we apply the Monod-Wyman-Changeux (MWC) model for coupled conformational changes of protein clusters~\cite{changeuxPNAS67}. Proteins are permitted to adopt either an R state (``$S_i=+1$'') or a T state (``$S_i=-1$Ó), and the energy cost for a given choice $\{S_i\}$ of the states of a one-dimensional cluster of correlated proteins can be expressed as
\begin{equation}
\label{energyMWC}
\mathcal{H}=NE_R-\frac{\Delta E_{RT}}{2} \sum_{i=1}^N\left(S_i-1\right)-J\sum_{i=1}^NS_iS_{i+1}  \ .
\end{equation}
Here, $J$ is the energy scale for the \emph{cooperativity} between adjacent units, $E_R$ is the binding energy of the IA unit to the substrate in the R state, and $E_R+\Delta E_{RT}$ in the T state. The difference between the T and R binding energies depends linearly on the unit tension $f$ as $\Delta E_{RT}(f)=\Delta E_0(1-f/f_{RT})$, with $\Delta E_0$ the zero tension energy difference and $\delta =\Delta E_0/f_{TR}$ the change in size of the IA unit due to the conformational transition. The MWC model maps onto the one-dimensional Ising model, which means that we can directly obtain the (grand-canonical) partition function
\begin{equation}
\label{partitionfct}
\begin{split}
\mathcal{Z} & = \sum_{N=1}^{\infty} e^{\beta \mu N} \sum_{\{S_i\}_{1=1,N}} e^{-\beta \mathcal{H}}  \\
& = \frac{1}{\displaystyle 1-e^{ -\beta(\epsilon_{+}-\tilde{\mu})}}+\frac{1}{\displaystyle 1-e^{-\beta(\epsilon_{-}-\tilde{\mu})}}    \ .
\end{split}
\end{equation}
The quantity $\tilde{\mu}=\mu-E_R+J$  plays the role of the effective integrin chemical potential, and the energy scales $\epsilon_{\pm}$ depend on the external tension as
\begin{equation}
\label{epsilon}
\begin{split}
\epsilon_{\pm}(f) & = \frac{\Delta E_{RT}(f)}{2} - k_BT \Bigg[ \cosh\left(\frac{\Delta E_{RT}(f)}{2k_BT}\right) \\
& \pm \sqrt{ \sinh^2 \left(\frac{\Delta E_{RT}(f)}{2k_BT}\right) +e^{-4\beta J}} \  \Bigg]   \ .
\end{split}
\end{equation}
If we neglect the weak, logarithmic dependence of the substrate friction coefficient on $N$ (see Eq.~(\ref{fiber/substrate})), we obtain our final expression for the thermally averaged total tension $\langle T \rangle = \langle N \rangle f(\zeta)$ along the stress fiber
\begin{equation}
\label{tension}
\langle T \rangle \approx \frac{1}{\displaystyle 1-e^{ -\beta(\epsilon_{+}-\tilde{\mu})}}f(\zeta)   \ ,
\end{equation}
where $\epsilon_{+}$  depends on $\zeta$ through Eqs.~(\ref{force/friction}) and (\ref{epsilon}).

\begin{figure}
\centering
\includegraphics{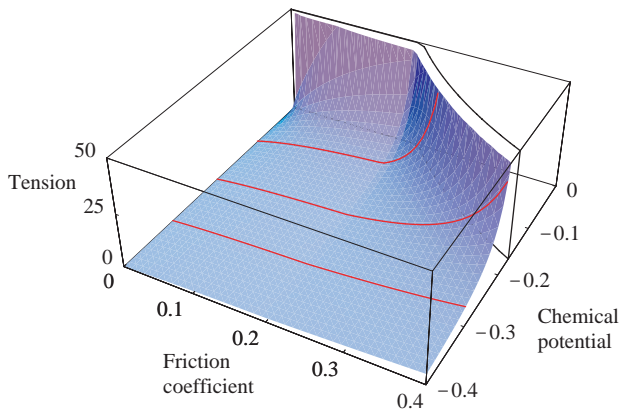} 
\caption{Average tension $\langle T \rangle$  (in units of  $f_0$) as a function of the chemical potential $\tilde{\mu}$  (in unit of $k_BT$) and the friction coefficient  $\zeta$ (in units of  $\zeta_0$), for parameters  $k=5$,  $\Delta E_0=0.5k_BT$, $f_{TR}=0.1f_0$  and $J=10k_BT$  (see text). The full curves show the dependence of the tension on the friction for fixed chemical potential  $\tilde{\mu}/k_BT=-0.3$, $-0.2$ and $-0.1$, respectively.}
\label{fig2}
\end{figure}

Figure~\ref{fig2} shows the dependence of fiber tension $\langle T \rangle$ on substrate friction coefficient $\zeta$ and integrin chemical potential $\tilde{\mu}$. At low integrin chemical potentials, the clusters are small. The ``bare'' physico-chemical stress sensitivity of Eq.~(\ref{force/friction}) is the only mechanism at work, which provides only minimal stress detection. For intermediate chemical potentials, the tension is initially low and practically independent of substrate friction for low viscosities. However, the tension increases dramatically at a threshold value of the friction coefficient that depends linearly on chemical potential, due to a tension-induced condensation transition of the integrins. Finally, for large chemical potentials, the tension grows immediately when the substrate viscosity turns on.

The scenario for intermediate chemical potentials accounts qualitatively for the properties of focal adhesion stress-detection mechanism summarized in (i)-(iii). For what parameter values does it operate? Stress-detection requires the unit tension $f$ Ð of order 0.1 pN Ð to be comparable to the force-scale $f_{TR}$ for the conformational change. The adaptor size change $\delta$ should not exceed the size of a protein (of the order of a few nanometer), which means $\Delta E_0$ has to be of order 0.1 $k_BT$.  To ensure cooperativity, $J/(k_BT)$ must be large compared to one. 

The model thus predicts that inside the focal adhesion complex, adaptor proteins should be found that (i) are strongly coupled to neighbouring units, (ii) are linked to the stress fiber, (iii) undergo a large-scale conformational change between two nearly degenerate configurations. In conclusion, we have derived a simple model that describes the generation and regulation of the tension exerted by the focal adhesion on the ECM. This model is based on the idea of a ``slipping clutch'', where the cooperative, tension-induced aggregation of IA units causes the clutch to engage as the system crosses the threshold shown in Fig.~\ref{fig2}.

\acknowledgments
We would like to thank J.-F. Joanny and D. Riveline for useful discussions. The model was originally suggested by B. Geiger at the Munich 2000 Biophysics Conference.


\begin{thebibliography}{99}

\bibitem{howardbook}
J. Howard, \emph{Mechanics of Motor Proteins and the Cytoskeleton} (Sinauer Associates Inc., 2001).

\bibitem{geigerCOCB01}
B. Geiger and A. Bershadsky, Curr. Opin. Cell Biol. {\bf 13}, 584 (2001).

\bibitem{wehrleTCB02}
B. Wehrle-Haller and B.A. Imhof, Trends Cell Biol. {\bf 12}, 382 (2002).

\bibitem{braybook}
D. Bray, \emph{Cell Movements: From Molecules to Motility}, 2$^{nd}$ Edition  (Garland Publishing, NY, 2001).

\bibitem{joannyPRL03}
J.-F. Joanny, F. J\"ulicher, and J. Prost, Phys. Rev. Lett. {\bf 90}, 168102 (2003).

\bibitem{pelhamPNAS97}
R.J. Pelham and Y.-L. Wang, Proc. Natl. Acad. Sci. USA {\bf 94}, 13661 (1997).

\bibitem{beningoTCB02}
K.A. Beningo and Y.-L. Wang, Trends Cell Biol. {\bf 12}, 79 (2002).

\bibitem{choquetCell97}
D. Choquet, D.P. Felsenfeld, and M.P. Sheetz, Cell {\bf 88}, 39 (1997).

\bibitem{tanPNAS03}
J.L. Tan, J. Tien, D.M. Pirone, D.S. Gray, K. Bhadriraju, and C.S. Chen, Proc. Natl. Acad. Sci. USA {\bf 100}, 1484 (2003).

\bibitem{loBPJ00}
C.-M. Lo, H.-B. Wang, M. Dembo, and Y.-L. Wang, Biophys. J. {\bf 79}, 144 (2000).

\bibitem{balabanNCB01}
N.Q. Balaban, U.S. Schwarz, D. Riveline, P. Goichberg, G. Tzur, I. Sabanay, D. Mahalu, S. Safran, A. Bershadsky, L. Addadi, and B. Geiger, Nature Cell Biol. {\bf 3}, 466 (2001).

\bibitem{schwartzBPJ02}
U.S. Schwarz, N.Q. Balaban, D. Riveline, A. Bershadsky, B. Geiger, and S.A. Safran, Biophys. J. {\bf 83}, 1380 (2002).

\bibitem{smilenovSci98}
L.B. Smilenov, A. Mikhailov, R.J. Pelham, E.E. Marcantonio, and G.G. Gundersen, Science {\bf 286}, 1172 (1999).

\bibitem{suterJCB98}
D.M. Suter, L.D. Errante, V. Belotserkovsky, and P. Forsher, J. Cell Biol. {\bf 141}, 227 (1998).

\bibitem{rivelineJCB01}
D. Riveline, E. Zamir, N.Q. Balaban, U.S. Schwarz, T. Ishizaki, S. Narumiya, Z. Kam, B. Geiger, and D. Bershadsky, J. Cell Biol. {\bf 153}, 1175 (2001).

\bibitem{chicurelCOCB98}
M.E. Chicurel, C.S. Chen, and D.E. Ingber, Curr. Opin. Cell Biol. {\bf 10}, 232 (1998).

\bibitem{guttenbergLangmuir00}
Z. Guttenberg, A.R. Bausch, B. Hu, R. Bruinsma, L. Moroder, and E. Sackmann, Langmuir {\bf 16}, 8984 (2000).

\bibitem{komuraEPJE00}
S. Komura and D. Andelman, Eur. Phys. J. E {\bf 3}, 259 (2000).

\bibitem{doibook}
M. Doi and S.F. Edwards, \emph{The Theory of Polymer Dynamics} (Clarendon Press, Oxford, 1986).

\bibitem{changeuxPNAS67}
J.-P. Changeux, J. Thi\'ery, Y. Tung, and C. Kittel, Proc. Natl. Acad. Sci. USA {\bf 57}, 335 (1967).



\end{thebibliography}
\end{document}